\documentclass[12pt,a4paper]{article}

\usepackage{amssymb}
\usepackage{graphics}
\usepackage[dvips]{epsfig}
\usepackage{natbib}
\bibpunct{(}{)}{;}{a}{}{,}

\begin{document}  

\title{Nucleosynthetic signatures of the first stars}
\date{March 2005}
\maketitle

Anna Frebel$^{1}$, Wako Aoki$^{2}$, Norbert Christlieb$^{2,3}$, Hiroyasu
Ando$^{2}$, Martin Asplund$^{1}$, Paul S. Barklem$^{4}$, Timothy
C. Beers$^{5}$, Kjell Eriksson$^{4}$, Cora Fechner$^{3}$, Masayuki
Y. Fujimoto$^{6}$, Satoshi Honda$^{2}$, Toshitaka Kajino$^{2}$, Takeo
Minezaki$^{7}$, Ken`ichi Nomoto$^{8}$, John E. Norris$^{1}$, Sean
G. Ryan$^{9}$, Masahide Takada-Hidai$^{10}$, Stelios Tsangarides$^{9}$, Yuzuru
Yoshii$^{7}$
\\
\\
$^{1}$ Research School of Astronomy \& Astrophysics, The Australian National
  University, Cotter Road, Weston Creek ACT 2611, Australia 
\\
$^{2}$ National Astronomical Observatory of Japan, 2-21-1, Osawa, Mitaka,
Tokyo 181-8588, Japan
\\
$^{3}$ Hamburger Sternwarte, Gojenbergsweg 112, 21029 Hamburg, Germany
\\
$^{4}$ Department of Physics \& Space Sciences, Uppsala Astronomical
Observatory, Box 515, \mbox{SE-751} 20 Uppsala, Sweden
\\
$^{5}$ Department of Physics \& Astronomy, and JINA: Joint Institute for
Nuclear Astrophysics, Michigan State University, East Lansing, MI 48824, USA
\\
$^{6}$ Department of Physics, Hokkaido University, Sapporo, Hokkaido 060-0810,
Japan
\\
$^{7}$ Institute of Astronomy, School of Science, University of Tokyo,
Mitaka, Tokyo 181-0015, Japan
\\
$^{8}$ Department of Astronomy, School of Science, University of Tokyo, Tokyo
113-0033, Japan
\\
$^{9}$ Department of Physics and Astronomy, Open University, Walton Hall,
 Milton Keynes MK76AA, England, UK
\\
$^{10}$ Liberal Arts Education Center, Tokai University, 1117 Kitakaname,
Hiratsuka-shi, Kanagawa 259-1292, Japan
\\
\\

\textbf{The chemically most primitive stars provide constraints on the nature
of the first stellar objects that formed in the Universe; elements other than
hydrogen, helium and traces of lithium within these objects were generated by
nucleosynthesis in the very first stars. The relative abundances of elements
in the surviving primitive stars reflect the masses of the first stars,
because the pathways of nucleosynthesis are quite sensitive to stellar
masses. Several models$^{1-5}$ have been suggested to explain the origin of
the abundance pattern of the giant star HE~0107$-$5240, which hitherto
exhibited the highest deficiency of heavy elements known.$^{1,6}$ Here we
report the discovery of HE~1327$-$2326, a subgiant or main-sequence star with
an iron abundance about a factor of two lower than that of HE~0107$-$5240. Both
stars show extreme overabundances of carbon and nitrogen with respect to iron,
suggesting a similar origin of the abundance patterns. The unexpectedly low Li
and high Sr abundances of HE~1327$-$2326, however, challenge existing
theoretical understanding: none predicts the high Sr abundance or provides a
Li depletion mechanism consistent with data available for the most metal-poor
stars.}  
\\ 
\\ 
The star HE~1327$-$2326 (coordinates right ascension,
$RA=13$\,h $30$\,m $06$\,s and declination $\delta=-23^{\circ}\;41'\;51''$ at
Equinox 2000; apparent visual magnitude $V=13.5$\,mag) was found among $1777$
bright $(10<B<14)$ and apparently metal-poor objects with partly saturated
spectra selected from the Hamburg/ESO objective-prism survey (HES)$^{7}$. Its
extreme metal-weakness was first appreciated in a medium-resolution
($\delta=0.5$\,nm) follow-up spectrum obtained in April 2003 at the $3.6$\,m
telescope of the European Southern Observatory, Chile. In May and June 2004,
observations with the High Dispersion Spectrograph$^{8}$ at the Subaru
Telescope, Hawaii were taken. Figure 1 shows a portion of this spectrum and
gives more details on the discovery.

\begin{figure}
 \begin{center} \epsfig{file=frebelFig1_latex.ps, clip=,
  width=12cm, bbllx=35, bblly=130, bburx=527, bbury=712}
  \caption{Comparison of high-resolution spectra of HE~1327$-$2326 with G64-12
and CS 22876-032. The latter is a double-lined spectroscopic binary. All three
stars have similar effective temperature and gravity. In HE~1327$-$2326 note
the absence of the FeI line at $393$\,nm together with the appearance of CH
lines. Our best fit for the Ca\,II\,K ($\lambda= 393.4$\,nm) and CH band
abundances is plotted in red. Additional CH fits (in red) for our carbon
abundance \mbox{[C/Fe]}$ = 4.1 \pm 0.2$\,dex are shown as well. The inset
shows our strongest detected Fe I line ($0.64$\,pm equivalent width) at
$386$\,nm in the Subaru data from which we derive
$\mbox{[Fe/H]}_{\mbox{\scriptsize non-LTE}} = -5.4\pm0.2$. From our
medium-resolution spectrum we initially estimated \mbox{[Fe/H]}$ = -4.0$ for
HE~1327$-$2326 using the apparent strength of the Ca\,II\,K line index KP of
Beers et al.$^{21}$ and an approximate colour. The high-resolution data
revealed the presence of strong interstellar Ca\,II\,K which had not been
resolved at the lower resolution. See the prominent feature blueward of the Ca
II\,K\,line. The interstellar Ca is consistent with a colour excess of $E(B-V)
= 0.08$ along the line-of-sight to HE~1327$-$2326. The presence of CH lines
near the Ca\,II\,K line in HE~0107$-$5240 caused a similar problem in the
initial recognition of its extreme metal-deficiency. These effects should be
taken into account in future searches for more stars with similar heavy
element deficiency. Our high-resolution (resolving power $R =
\Delta\lambda/\lambda = 50,000$), high signal-to-noise ratio ($S/N = 160$ per
pixel at $400$\,nm) data covers the wavelength range of $304$ to $674$\,nm.}
\end{center}
\end{figure}

An important difference between HE~0107$-$5240 and HE~1327$-$2326 is their
evolutionary status. HE~0107$-$5240 is a giant and has evolved off the
main-sequence and up the red giant branch. Consequently, internal mixing has
possibly dredged up processed material (e.g. C, N) from the stellar interior
to the surface and altered the abundances of this object. In contrast,
HE~1327$-$2326 is relatively unevolved and located on either the main-sequence
or the subgiant branch.

We use model atmosphere techniques and the assumption of local thermodynamical
equilibrium (LTE) to perform our chemical abundance analysis with non-LTE
corrections where available. There are no significant differences between the
subgiant and the main-sequence solution. See Table 1 for more details on the
analysis procedure and the derived abundances.

\begin{table}
{
\caption{Abundance ratios of HE~1327$-$2326. Using colour-effective
temperature ($T_{\mbox{\scriptsize eff}}$) relations$^{24}$, we determine an
effective temperature of $T_{\mbox{\scriptsize eff}}=6180\pm80$\,K from
broad-band $UBVRI$ photometry obtained with the MAGNUM telescope$^{25}$,
Hawaii, and $JHK$ magnitudes taken from the Two Micron All Sky
Survey$^{26}$. To constrain the gravity ($\log g$), we use the proper
motion$^{27}$ ($\mu=0.073 \pm0.006$\,arcsec/yr) of HE~1327$-$2326 to set
limits on its distance (assuming that the Galactic escape velocity is
$500$\,km\,s$^{-1}$ and larger than the transverse velocity). It follows that
the absolute $V$ magnitude is $M_{V}>2.7$\,mag which constrains the
evolutionary status of the star and therefore $\log g$. Inspection of a
12-billion-year isochrone with $\mbox{[Fe/H]} = -3.5$ (the most metal-poor one
available$^{28}$) results in two solutions for the surface gravity $\log g$:
$3.7$ and $4.5$ in cgs units. Owing to weak observational constraints on $\log
g$, it is currently not possible to determine which of these solutions is
correct. $T_{\mbox{\scriptsize eff}}$ and $\log g$ are the input for two
different plane-parallel LTE model atmospheres we use to derive the abundances
for both gravity solutions; a MARCS model (Gustafsson, B. et al. 2005,
manuscript in preparation) tailored for the chemical composition of
HE~1327$-$2326, and a Kurucz model$^{29}$. Both sets of abundances agree to
within $0.1$\,dex. Using spectrum synthesis we determine the C and N
abundances from CH and NH molecular bands and the upper limit for O from OH
molecular bands in the ultra-violet spectral range. Apart from the molecular
features, only 13 weak absorption lines are detected in spectral regions not
affected by the wings of the Balmer lines and thus suitable for our abundance
analysis. Solar abundances were taken from ref.30. Typical statistical
$1\,\sigma$ errors in our abundances are $0.2$\,dex. Possible systematic
errors are judged to be of the same order of magnitude. Non-LTE corrections
are included where available$^{22}$. For the calculations of the non-LTE
abundance ratios, the non-LTE iron abundance has been used.}

\begin{center}
\begin{tabular}{lrr}\hline\hline
\rule{-.8ex}{2.3ex}
Element &\multicolumn{2}{r}\mbox{[Element/Fe]}, A(Li), \mbox{[Fe/H]}\\
        &Subgiant        &Dwarf\\\hline\rule{-.8ex}{2.3ex} 
Li          & $<1.6       $ & $<1.6       $\\
C           & $4.1 \pm 0.2$ & $3.9 \pm 0.2$ \\
N           & $4.5 \pm 0.2$ & $4.2 \pm 0.2$ \\
O           & $<4.0       $ & $<3.7      $  \\
Na (LTE)    & $2.4 \pm 0.2$ & $2.4 \pm 0.2$ \\
Na (non-LTE)& $2.0 \pm 0.2$ & $2.0 \pm 0.2$ \\
Mg (LTE)    & $1.7 \pm 0.2$ & $1.7 \pm 0.2$ \\
Mg (non-LTE)& $1.6 \pm 0.2$ & $1.6 \pm 0.2$ \\
Al (LTE)    & $1.3 \pm 0.2$ & $1.3 \pm 0.2$ \\
Al (non-LTE)& $1.7 \pm 0.2$ & $1.7 \pm 0.2$ \\
Ca I        & $0.1 \pm 0.2$ & $0.1 \pm 0.2$ \\
Ca II       & $0.9 \pm 0.2$ & $0.8 \pm 0.2$ \\
Ti          & $0.6 \pm 0.2$ & $0.8 \pm 0.2$ \\
Fe (LTE)    & $-5.6 \pm 0.2$& $-5.7 \pm0.2$ \\
Fe (non-LTE)& $-5.4 \pm 0.2$& $-5.5 \pm0.2$ \\
Sr (LTE)    & $1.0 \pm 0.2$ & $1.2 \pm 0.2$ \\
Sr (non-LTE)& $1.1 \pm 0.2 $& $1.3 \pm 0.2$ \\
Ba (LTE)    & $<1.4	   $& $<1.7      $  \\	
Ba (non-LTE)& $<1.4       $ & $<1.7$        \\	 \hline
   \end{tabular}
\end{center}
}
\end{table}

HE~1327$-$2326 sets a new record for the star with the lowest iron abundance:
\mbox{[Fe/H]}$_{\mbox{\scriptsize non-LTE}}=-5.4 \pm 0.2$ (where \mbox{[A/B]}$
= \log(N_{A}/N_{B}) - \log(N_{A}/N_{B})_\odot$ for the number N of atoms of
element A and B and $\odot$ refers to the Sun), derived from four detected
weak Fe I lines and corrected for non-LTE. This low value corresponds to
$1/250,000$ of the solar Fe value. To illustrate the nature of such extremely
Fe-deficient stars, in Figure 2 we compare our abundances with those of
HE~0107$-$52406 (\mbox{[Fe/H]}$_{\mbox{\scriptsize non-LTE}} = -5.2$,
refs. 1,6 and 9), the only other star known to have \mbox{[Fe/H]}$<-4.0$. Both
objects have very large overabundances of C and N relative to Fe. Although
both stars also share a deficiency of Ca and Ti as well as Fe, the Mg and Al
abundances relative to Fe in HE~1327$-$2326 are more than one order of
magnitude higher than those of HE~0107$-$5240.

Surprisingly, we do not detect the Li I doublet at $670.7$\,nm in
HE~1327$-$2326, although it is found in similarly unevolved metal-poor stars
that permit observational determination of the primordial Li abundance and the
associated baryon density of the Universe.  Our upper limit for Li of
A(Li)$=1.6$ (where A(\mbox{Li})$=\log(N(\mbox{Li})/N(\mbox{H}))+12.00)$ is
significantly lower than the primordial Li value of A(Li)$=2.62\pm0.05$,
inferred from the baryon density estimated by the Wilkinson Microwave
Anisotropy Probe together with standard Big Bang nucleosynthesis.$^{10}$ It is
even below A(Li)$=2.0$, found by Ryan et al.$^{11}$ at
\mbox{[Fe/H]}$=-3.5$. This apparent absence does not result from the
evolutionary status of HE~1327$-$2326, because in more evolved metal-poor
subgiants such as HD140283, Li has not been depleted. Li weakness occurs
rarely in metal-poor stars near the main sequence turn-off: at effective
temperature $T_{\mbox{\scriptsize eff}}=6200$\,K only $1$ in $\sim20$ shows
this effect.$^{11}$ In HE~1327$-$2326, the weak lines show no evidence for
rotational broadening and offer no support for the ultra-Li depleted rapid
rotator model$^{12}$, which assumes membership of a binary system, or for
rotationally induced mixing in general.$^{13}$ Other explanations might
include gravitational settling$^{14}$ or diffusion, but these have not
operated efficiently in other metal-deficient stars of similar temperature and
gravity.

\begin{figure}
 \begin{center} \epsfig{file=frebelFig2_updated.ps, clip=,
  width=17cm, bbllx=63, bblly=425, bburx=481, bbury=712}
  \caption{Abundance patterns of HE~1327$-$2326 (subgiant solution, filled
circles) and HE~0107$-$5240 (open squares). Typical $1\,\sigma$ errors of
$0.2$\,dex are shown in the plot. Upper limits are indicated by an arrow. We
adopt the same non-LTE correction for both stars$^{22}$, which lead to the
modification of the published abundances of HE~0107$-$5240. Consequently, we
adopt \mbox{[Fe/H]}$_{\mbox{\scriptsize non-LTE}} = -5.2$ (ref. 9) as the iron
abundance of HE~0107$-$5240. These two most Fe-poor stars both have very large
C enhancement relative to Fe by a factor of $\sim5,000$ (HE~0107$-$5240) and
$\sim10,000$ (HE~1327$-$2326). N/Fe is $\sim30,000$ times the solar value
(considering the subgiant solution) in HE~1327$-$2326 whereas it is $\sim200$
times the solar value in HE~0107$-$5240. The upper limit for the O abundance
of HE~1327$-$2326 is $\mbox{[O/Fe]} < 4.0$. Oxygen in HE~0107$-$5240 has
recently been determined$^{23}$ to be $\mbox{[O/Fe]} = 2.3$ which is of the
same order as its \mbox{[N/Fe]} value. These enormous overabundances in CNO
elements suggest that both stars belong to a group of objects sharing a common
formation scenario. HE~1327$-$2326 and HE~0107$-$5240 have Ca/Fe and Ti/Fe
abundance ratios which are enhanced by factors less than 10 compared to the
Sun. The light element ratios Na/Fe, Mg/Fe and Al/Fe, as well as,
surprisingly, Sr/Fe, are all enhanced by factors of $\sim10$ to $\sim100$ in
HE~1327$-$2326. Of these four elements only Na and Mg are detected in
HE~0107$-$5240 with element/Fe ratios close to the solar value. As for several
other elements an upper limit for Ba has been measured in both stars. The
Sr/Ba ratio is crucial to identify the origin for the Sr and other heavy
elements.}
\end{center}
\end{figure}

Another surprising result is the high Sr abundance (see Figure 2). Our
observed lower limit for the Sr/Ba ratio (\mbox{[Sr/Ba]}$_{\mbox{\scriptsize
non-LTE}}>-0.4$) suggests that Sr was not produced in the main ``slow''
neutron-capture (s-) process occurring in asymptotic giant branch (AGB) stars,
given that the typical ratio in Fe-deficient, C-rich and s-process enhanced
stars is$^{15}$ \mbox{[Sr/Ba]}$_{\mbox{\scriptsize non-LTE}}<-1.0$. The weak
s-process occurring in formerly more massive stars ($M>10\,M_{\odot}$ might
yield Sr with little or no Ba, although theoretical calculations suggest that
this process is inefficient at low metallicity.$^{16}$ Alternatively, the main
``rapid'' neutron-capture (r-) process might have produced the Sr, given that
\mbox{[Sr/Ba]}$_{\mbox{\scriptsize non-LTE}}>-0.4$ is consistent with the
ratio observed in metal-poor, r-process enhanced stars, in which$^{17}$
\mbox{[Sr/Ba]}$_{\mbox{\scriptsize non-LTE}}\sim-0.3$ (non-LTE corrections
were applied as for HE~1327$-$2326). It is therefore possible that the Sr was
formed in a type II supernova (SNII), currently believed to be a site of the
main r-process, that expelled material from which HE~1327$-$2326
formed. Another possibility is the recently proposed process that might have
provided light neutron-capture elements in the early Galaxy.$^{16}$ Although
the site and details of this newly suggested process are not yet known, it
might provide an explanation for the excesses of light neutron-capture
elements, including Sr, seen in very metal-poor stars. To discriminate the
contribution of such processes, further constraints on the Ba abundance, as
well as further information on other neutron-capture elements, from higher
quality data, are needed.

With its distinctive chemical composition, HE~1327$-$2326 is an ideal tool for
``stellar archaeology'' and a better understanding of the early enrichment
history, and the first mass function of stars in the Universe. Although two
stars, HE~1327$-$2326 and HE~0107$-$5240, are now known with
\mbox{[Fe/H]}$<-5.0$, none has been found in the range
$-5<\mbox{[Fe/H]}<-4$. This is suggestive of an underlying, yet to be
explored, fundamental physical mechanism in the formation of the first stellar
objects$^{2}$ which may differ from that for stars more metal-rich than
\mbox{[Fe/H]}$= -4.0$.

HE~1327$-$2326 permits exploration of the many formation theories of the
earliest stars proposed for HE~0107$-$5240. One idea is the pre-enrichment of
the gas clouds from which the two stars formed by a first generation ``faint''
type-II supernova ($M \sim 25\,M_{\odot}$) experiencing mixing and
fallback$^{3}$, or rotating, massive objects ($M\sim200-300\,M_{\odot}$,
ref.18) In this case, HE~1327$-$2326 would be an early Population II star,
formed from gas enriched by one (or possibly a few) of the first SNII. In
particular, the quantitative predictions of abundance patterns by updated
models of ``faint'' SNII (N. Iwamoto, H. Umeda, N. Tominaga, K. Maeda, private
communication) agree with the chemical abundance patterns of
HE~1327$-$2326. The diversity of Mg/Fe ratio found in HE~1327$-$2326 and
HE~0107$-$5240 is easily explained by the variety of mixing and fallback. A
remaining problem is the Sr whose production is not yet included in these
type-II supernova models.

Another idea is the accretion of heavy elements from the ISM onto first
generation low-mass stars (Population III).$^{19,1,2}$ In this case, however,
the high abundances of lighter elements must be explained by other
processes. The unevolved status of HE~1327$-$2326, the scenario in which
internal mixing processes lead to the dredge-up of processed material can be
excluded. A remaining possibility is mass transfer from an AGB star within a
binary system.$^{1,5}$ Mass transfer could naturally account for the Li
depletion$^{20}$ found in HE~1327$-$2326. The high Sr abundance is
problematic, along with the non-detection of Ba, which can not be explained by
the s-process expected to occur in AGB stars. A crucial test for this scenario
is to check the binarity of this object, for which long period radial velocity
monitoring is required.

\begin{enumerate}
\item[1.] Christlieb, N. et al. A stellar relic from the early Milky Way. Nature 419,
   904-906 (2002)

\item[2.] Shigeyama, T., Tsujimoto, T. \& Yoshii, Y. Excavation of the First
   Stars. Astrophys. J. 568, L57-L60 (2003)

\item[3.] Umeda, H. \& Nomoto, K. First-generation black-hole-forming
   supernovae and the metal abundance pattern of a very iron-poor star. Nature
   422, 871-873 (2003)

\item[4.] Limongi, M., Chieffi, A. \& Bonifacio, P. On the Origin of
   HE~0107-5240, the Most Iron-deficient Star Presently
   Known. Astrophys. J. 594, L123-L126 (2003)

\item[5.] Suda, T., Aikawa, M., Machida, M. N., Fujimoto, M. Y. \& Iben, I. Jr. Is HE~
   0107-5240 A Primordial Star? The Characteristics of Extremely Metal-Poor
   Carbon-Rich Stars. Astrophys. J. 611, 476-493 (2004)

\item[6.] Christlieb, N. et al. HE~ 0107-5240, a Chemically Ancient Star. I. A
   Detailed Abundance Analysis. Astrophys. J. 603, 708-728 (2004)

\item[7.] Wisotzki, L. et al. The Hamburg/ESO survey for bright QSOs. III. A
   large flux-limited sample of QSOs. Astron. Astrophys. 358, 77-87 (2000)

\item[8.] Noguchi, K. et al. High Dispersion Spectrograph (HDS) for the Subaru
   Telescope. Publ. Astron. Soc. Jap., 54, 855-864 (2002)

\item[9.] Beers, T. C. \& Christlieb, N. The Discovey and Analysis of Very
Metal-Poor Stars in The Galaxy. Annu. Rev. Astron. Astroph, in press (2005)

\item[10.] Coc, A., Vangioni-Flam, E., Descouvemont, P., Adahchour, A.,
Angulo, C. Updated Big Bang Nucleosynthesis Compared with Wilkinson Microwave
Anisotropy Probe Observations and the Abundance of Light
Elements. Astrophys. J. 600, 544-552 (2004)

\item[11.] Ryan, S. G., Norris, J. E., Beers, T. C. The Spite Lithium Plateau:
Ultrathin but Postprimordial. Astrophys. J. 523, 654-677 (1999)

\item[12.] Ryan, S. G., Gregory, S. G., Kolb, U., Beers, T. C. \& Kajino,
    T. Rapid Rotation of Ultra-Li-depleted Halo Stars and Their Association
    with Blue Stragglers. Astrophys. J. 571, 501-511 (2002)

\item[13.] Pinsonneault, M. H.; Walker, T. P.; Steigman, G. \& Narayanan,
V. K. Halo Star Lithium Depletion. Astrophys. J. 527, 180-198 (1999)

\item[14.] Richard, O., Michaud, G. \& Richer, J. Models of Metal-poor Stars
with Gravitational Settling and Radiative Accelerations. III. Metallicity
Dependence. Astrophys. J. 580, 1100-1117 (2002)

\item[15.] Aoki, W., Norris, J. E., Ryan, S. G., Beers, T. C. \& Ando,
H. Detection of Lead in the Carbon-rich, Very Metal-poor Star LP 625-44: A
Strong Constraint on s-Process Nucleosynthesis at Low
Metallicity. Astrophys. J. 536, L97-L100 (2000)

\item[16.] Travaglio, C., Gallino, R., Arnone, E., Cowan, J., Jordan, F. \&
    Sneden, C. Galactic Evolution of Sr, Y, and Zr. A Multiplicity of
    Nucleosynthetic Processes. Astrophys. J. 601, 864-884 (2004)

\item[17.] Christlieb, N. et al. The Hamburg/ESO R-process Enhanced Star
survey (HERES). I. Project description, and discovery of two stars with strong
enhancements of neutron-capture elements. Astron. Astrophys. 428, 1027-1037
(2004)

\item[18.] Fryer, C. L., Woosley, S. E., Heger, A. Pair Instability
Supernovae, Gravity Waves, and Gamma-Ray Transients. Astrophys. J. 550,
372-382 (2001)

\item[19.] Yoshii, Y. Metal Enrichment in the Atmospheres of Extremely
Metal-Deficient Dwarf Stars by Accretion of Interstellar
Matter. Astron. Astrophys. 97, 280-290 (1981)

\item[20.] Norris, J. E., Ryan, S. G., Beers, T. C. \& Deliyannis,
C. P. Extremely Metal-Poor Stars. III. The Li-depleted Main-Sequence Turnoff
Dwarfs. Astrophys. J. 485, 370-379 (1997)

\item[21.] Beers, T. C. Rossi, S., Norris, J. E., Ryan, S. G. \& Shefler,
T. Estimation of Stellar Metal Abundance. II. A Recalibration of the Ca II K
Technique, and the Autocorrelation Function Method. Astron. J. 117, 981-1009
(1999)

\item[22.] Asplund, M. New light on stellar abundances analyses: departures
from LTE and homogeneity. Annu. Rev. Astron. Astroph, in press (2005)

\item[23.] Bessell, M. S., Christlieb, N. \& Gustafsson, B. On the Oxygen
Abundance of HE 0107-5240. Astrophys. J. 612, L61-L63 (2004)

\item[24.] Alonso, A., Arribas, S. \& Martinez-Roger, C. The empirical scale
of temperatures of the low main sequence (F0V-K5V). Astron. Astrophys. 313,
873-890 (1996)

\item[25.] Yoshii, Y. The MAGNUM Project: AGN Variability as a New Technique
    for Distance Determination. New Trends in Theoretical and Observational
    Cosmology. K. Sato \& T. Shiromizu (editors), Universal Academy, Tokyo,
    235-244 (2002)

\item[26.] Cutri R. M. et al. 2MASS All-Sky Catalog of Point Sources
(Californian Institute of Technology, Pasadena, 2003);
http://irsa.ipac.caltech.edu/applications/Gator

\item[27.] Girard, T. M. et al. The Southern Proper Motion Program. III. A
    Near-Complete Catalog to V=17.5. 127, Astron. J. 127, 3060-3071 (2004)

\item[28.] Kim, Y., Demarque, P., Yi, S. K. \& Alexander, D. R. The Y2
Isochrones for Alpha-Element Enhanced Mixtures. Astrophys. J. Suppl. 143,
499-511 (2002)

\item[29.] Kurucz, R. L. Kurucz CD-ROM 13, ATLAS9 Stellar Atmosphere Programs
and 2 km/s Grid CD-ROM 13. (Smithonian Astrophysical Observatory, Cambridge,
1993); \\http://kurucz.harvard.edu/cdroms.html

\item[30.] Asplund, M., Grevesse, N. \& Sauval, A. J. The solar chemical
    composition. Cosmic Abundances As Records Of Stellar Evolution And
    Nucleosynthesis, F.N. Bash \& T.G Barnes (editors), ASP conference series
    (in press); preprint at \\http://www.arxiv.org/astro-ph/0410214 (2004)
\end{enumerate}

\textbf{Acknowledgements} We thank A. Steinhauer and C. Thom for obtaining
additional observations, N. Iwamoto, K. Maeda, T. Suda, N. Tominaga and
H. Umeda for valuable discussions and L. Wisotzki and D. Reimers for leading
the HES. This work was supported by the Astronomical Society of Australia
(A.F.), Australian Research Council (M.A., A.F., J.E.N.), Ministry of
Education, Culture, Sports, Science and Technology in Japan and JSPS (all
Japanese co-authors), Deutsche Forschungsgemeinschaft (N.C.), Swedish Research
Council (P.S.B., K.E.), US National Science Foundation (T.C.B.) and JINA
(T.C.B., N.C., A.F., J.E.N.). This work is based on data collected at the
Subaru Telescope, which is operated by the National Astronomical Observatory
of Japan.  
\\ 
\\ 
\textbf{Competing interests statement} The authors declare that they have no
competing financial interests.
\\ 
\\ 
\textbf{Correspondence} and requests for materials should be addressed to
A.F. \\(e-mail: anna@mso.anu.edu.au).

\end{document}